# Detection Latencies of Anomaly Detectors: An Overlooked Perspective?

Tommaso Puccetti[1], Andrea Ceccarelli[1]

**Affiliations**
1. Department of Mathematics and Informatics, University of Florence, Viale Morgagni 67/a, 50134 Firenze (FI), Italy.

## Abstract

The ever-evolving landscape of attacks, coupled with the growing complexity of ICT systems, makes crafting anomaly-based intrusion detectors (ID) and error detectors (ED) a difficult task: they must accurately detect attacks, and they should promptly perform detections. Although improving and comparing the detection capability is the focus of most research works, the timeliness of the detection is less considered and often insufficiently evaluated or discussed. In this paper, we argue the relevance of measuring the temporal latency of attacks and errors, and we propose an evaluation approach for detectors to ensure a pragmatic trade-off between correct and in-time detection. Briefly, the approach relates the false positive rate with the temporal latency of attacks and errors, and this ultimately leads to guidelines for configuring a detector. We apply our approach by evaluating different ED and ID solutions in two industrial cases: i) an embedded railway on-board system that optimizes public mobility, and ii) an edge device for the Industrial Internet of Things. Our results show that considering latency in addition to traditional metrics like the false positive rate, precision, and coverage gives an additional fundamental perspective on the actual performance of the detector and should be considered when assessing and configuring anomaly detectors.

## 1 Introduction

Information and Communication Technology (ICT) systems play a central role in daily operations, industrial processes, and even safety-critical contexts where a maximum frequency of failure per hour must be guaranteed, and attacks may have catastrophic consequences also impacting the well-being of people and the environment [1]. They are connected through the network, either to be accessible by the end-user or to orchestrate distributed components to provide an emergent behavior [2]. The complexity of these systems makes them susceptible to errors that may arise from diverse sources and are exposed to attacks.

Amongst the multiple approaches to tolerate attacks and failures, anomaly detection is a widely researched initiative. It consists of the deployment of anomaly-based error or intrusion detection systems [3], [4], [5], [6], [7], that monitor system indicators (often called *features*) during the execution of the system and utilize the values collected to suspect errors or attacks. The foundational assumption underlying the construction of these anomaly detectors is that errors and attacks produce discernible anomalies in the trend of the values collected from monitored features; luckily, this assumption often results realistic [8].

A promising solution to build the detector is training a machine learning model relying on a large set of features encoded as structured data. The detection of anomalies is then formulated as a binary classification problem in which we predict a label for any input data points provided to the detector (0 for normal, 1 for attack). The training process allows for the model to approximate the binary function that associates a label for each data point fed to the detector and can be performed either in a supervised or unsupervised way, depending on the availability of labels for the train datasets.

The exploitation of the temporal order between data points is a distinguishing feature of anomaly-based error detectors (EDs) and intrusion detectors (IDs). Incorporating the time dimension into the detector model enables the detection of contextual and group anomalies, in addition to punctual anomalies [12]. Contextual anomalies are data instances that are anomalous in a specific context but not otherwise, while group anomalies are collections of related data instances, anomalous to the entire data set [8]. If classification relies solely on the information conveyed by individual data points, the capabilities of the detector are limited to the detection of punctual anomalies, whose values are different from the modeled distribution. Time series data can be leveraged by models to make decisions based on the evolution of the system during the executions and provide larger power in describing the arising of anomalous behavior, triggered either by an attack or a system error. The literature is well aware of this, with numerous ID works that exploit the temporal order between data points [62], [63], [64], [65].

What we expect from an ID and ED, independent of the model, is to minimize the risk of false alarms to avoid triggering unnecessary responses, while being able to detect attacks or errors as soon as possible. In the literature, the correctness of the detection is a well-considered factor for evaluating the efficacy of ID and ED. It is measured in terms of how many data points in a sequence are classified correctly, by using state-of-the-art metrics for classification such as f1-score, precision, recall, and accuracy [9].

However, the time required to identify an occurring attack or failure, i.e., the attack and error latency, is often insufficiently considered when evaluating IDs and EDs, and successively for the configuration of the detector when deployed. Achieving high performance in terms of detection alone, without proper consideration of the time an attack or an error has been latent in the system, does not guarantee safety and security. Two main factors substantiate this lack of attention to latency, as can be reviewed from papers on the subject: the training dataset and the metrics used. First, while train datasets include timestamped data points, the way they are typically built is insufficient to model the evolution of an error or attack through time. They are composed of long periods of normal operativity, and long periods where multiple attacks are injected. This way, the transition between normal and anomalous behavior of the systems is reproduced only a few times in the data, preventing the detector from learning this dynamic. Second, widely accepted and typically used metrics like accuracy or f1-score count the number of correctly classified data points, but these metrics do not consider that multiple anomalous data points may belong to the same attack or failure event. In these cases, it may be not necessary to detect all the anomalous data points, but just one to start proper reaction strategies [10].

For these reasons, accuracy, f1-score, or similar metrics might seem satisfactory to benchmark and assess detector performances, but they may prevent a more sensitive configuration of the detector based on the operational context.

To overcome these limitations, we propose and apply an approach for evaluating IDs and EDs that consider latency and related metrics in addition to state-of-the-art classification metrics. This proves a complete understanding of a detector in a specific operational context, with the ability to define a trade-off between the correctness and timeliness of the detection.

Very practically, we assume that a system has operational requirements defined in terms of acceptable false positive rate, which may impact the availability of the system calling unnecessary response, and the need for rapid detection of attacks (to be detected before the system is compromised) and errors (to be detected before they escalate to a failure). Then, we proceed to evaluate our EDs and IDs as follows. Given an anomalous sequence, we set multiple false positive rate thresholds, and we compute the corresponding attack and error latency, which is the distance between the first anomalous data point and the first anomalous data point that is correctly classified. This gives us a way to identify the best configuration, based on the operational requirements of the system to defend.

We exercise the proposed approach and metric for evaluating IDS solutions executing two real systems: SPaCe, a cyber-physical system that orchestrates and optimizes public mobility, and Arancino, a smart board for distributed monitoring. Results show that the metric offers a new perspective on the intrusion detection problem and can be used to calibrate the IDS in a way that can be used efficiently in real deployment scenarios. The code to reproduce our results is available at [11].

The rest of the paper is organized as follows: Section 2 provides an overview of the background and related works concerning anomaly detection. In Section 3, we introduce our evaluation approach and discuss the motivation behind it. Section 4 elaborates on the methodological steps of our experimental campaign. Our results are presented in Section 5. We conclude our discussion in Section 6, where we summarize the practical takeaway from our work.

## 2 Background and Related Works

Section 2.1 reviews the notion of anomaly and how it is related to error and intrusion detection. Section 2.2 reviews recent and notorious machine learning solutions that have been applied to detect anomalies in IDs and EDs. Section 2.3 examines state-of-the-art metrics commonly used to evaluate ID and ED models, and differences from our approach. This includes a review of the most recent works that appeared in relevant conferences on the domain, in search of evaluation approaches and metrics adopted.

### 2.1 Anomalies

Anomalies are patterns in data that do not conform to a well-defined notion of normal behavior, that might be induced for various reasons such as malicious activity or system malfunctions [1]. In the case of EDs and IDs, anomalies can be observed by monitoring system *features*. Values collected from features are typically assembled in multivariate *data points*, that are processed by the detector at defined time instants.

It has been repeatedly proved that anomaly detection is a viable approach to detect occurring attacks and failures [3], [4], [5], as an alternative to signature-based detectors. The latter perform satisfactorily when aiming at observing known effects, for example, to recognize the signature of a known attack [6], [7]. However, signature-based detectors are weak against zero-day attacks or unprecedented events [13]. Especially in the case of zero days, these attacks exploit either new vulnerabilities or known vulnerabilities in novel ways and cannot be matched to known signatures. This problem cannot be neglected: zero-day attacks against a target system are a credible possibility, as confirmed by multiple reports in recent years [13], [14]. In anomaly-based detection, the problem of detecting attacks or errors is reduced to the problem of finding anomalies. However, anomaly-based detectors are still far from perfect [15], and more research is needed.

Usually, anomaly-based IDs and EDs rely on Machine Learning (ML) algorithms, where the anomaly detectors learn a model using a train set composed of both normal data and anomalous data. Most famous datasets in this domain are multivariate, with data points temporally ordered.

### 2.2 Anomaly Detection Solutions

Models can be classified into two categories, based on their ability to exploit the temporal context from the training dataset [8].

**Unordered data points.** Solutions belonging to this category do not benefit from the temporal dimension, i.e., there is no difference between shuffled data points or data points collected in a (temporal) sequence. The most researched approach is to use a machine learning-based detector that only relies on the features from a single data point $x_t$ and that is focused on detecting punctual anomalies. In this case, features related to the temporal or sequential order of data points are not exploited; on the opposite, they should be dropped when training to avoid acting as a label [16].

Amongst the most noticeable works, in [17] authors propose an ID based on XGBoost and principal component analysis (PCA). XGBoost is a supervised learning algorithm that combines the prediction of an ensemble of regression trees to output a final prediction [18]. The training dataset is transformed by applying a hybrid PCA-firefly algorithm with the purpose of dimensionality reduction. Other XGBoost-based solutions are proposed for example by [19], [20]. Random Forest is another popular ensemble-based solution that relies on supervised learning and bagging [21], [17], [20]. In particular, a random forest is a classifier that aggregates the decision of a decision tree classifier ensemble that is trained with a subset of the train set extracted by random sampling with replacement. Similarly, Extremely Randomized Trees [58] construct trees over every observation in the dataset but with different subsets of features.

Isolation Forest [22] is one of the most popular unsupervised approaches viable for intrusion detection that consists of explicitly isolating anomalies rather than profiling normal instances. The model composes an ensemble of isolation trees for a given data set. The anomalies are detected from those instances that have short average path lengths on the isolation trees. Other unsupervised approaches are K-Means [23], FastABOD [24], and LOF [25]. All these approaches are frequently used to evaluate IDs and EDs [15], [16], [20], [26], [27], [28], [29].

*In this paper we will use XGBoost, Extremely Randomized Trees, and Isolation Forest.*

**Exploit ordered data points.** Solutions belonging to this category are intended to exploit the temporal dimension or the sequence of data points, deriving knowledge from the variation of feature values in successive data points. There are two main approaches.

One approach is crafting new features that synthesize the changing of the features over consecutive data points. This way each data point $x_t$ encodes information about the previously observed values, allowing for contextual and collective anomaly detection. For example, if an existing feature is "percentage of RAM used", we may be interested in deriving features such as "difference in the percentage of RAM used concerning the previous observation" [30]. Automatic approaches to temporal feature crafting are widely explored in literature e.g., bag-of-features [31], complex temporal features [32], feature fusion [33], and distance-based features [34], which may be global, local, or embedded depending on the way they are computed.

The second approach is to use algorithms such as LSTM (Long-Short Term Memory [35]) that employ recurrent neural networks (RNN [36]) specifically crafted to consider the evolution of feature values. In this case, the approach is to predict the state $x_t$ of the data point at time $t$ (i.e., the values of the data point features observed at time $t$) using previous observations. This approach implies the definition of a threshold value that is compared to the difference between the predicted data point and the ground truth to decide if the data point is anomalous [37]. An example of this approach is the model in [60].

Amongst noticeable examples, the authors in [38] propose an LSTM-based anomaly detector to detect the abnormal behavior of the controller area network (CAN) bus under tampering attacks. A similar approach is applied in [39] to build an LSTM model which is combined with a machine learning algorithm to detect jump anomalies in financial time series.

Many approaches rely on LSTM to extract temporal features and on autoencoders to estimate the input time sequence. In this case, the loss between the ground truth sequence and the estimated one is compared to a threshold value: if the loss exceeds the threshold the sequence is considered anomalous [40], [41], [42], [43]. Authors in [61] combine a convolutional network to extract relevant features from the input and LSTM technology for sequence classification tasks (LSTMCD) [61].

Amongst noticeable examples, the authors in [44] apply a deep learning-based anomaly detection approach (DeepAnT) for time series data, including multivariate data. DeepAnT relies on unsupervised training and unlabeled data to capture and learn the data distribution that is used to forecast the normal behavior of a time series.

### 2.3 ID Metrics and Evaluation

Anomaly Detectors are typically evaluated based on state-of-the-art metrics for classification. This approach is applied regardless of the system, the type of anomalies targeted, and the possibility of taking advantage of temporal features. The most used approach measures how many data points are correctly classified in terms of True Positives (TP), True Negatives (TN), False Positives (FP), and False Negatives (FN).

Different metrics can be derived from the above. *Accuracy* measures the correct predictions in a classification task,

and it is defined as A = (TP + TN) / (TP + TN + FP + FN). Accuracy is an excellent summarizing metric, but it is misleading in the case of unbalanced datasets when the number of normal data points differs significantly from the number of anomalous data points. In this case, it should be replaced or complemented by other metrics.

Other notorious metrics focus on the detection capability and do not include TNs. Recall (also called coverage) measures the fraction of anomalies that are detected, and it is measured as R = TP / (TP + FN). Precision measures the fraction of detection that are anomalies, and it is measured as P = TP / (TP + FP). Precision and recall can be used to plot the precision-recall curve, which calculates the trade-off between precision and recall for different detection thresholds.

Still related to precision and recall, the f1-score is the harmonic mean of the precision and recall F1 = 2 × ((P × R) / (P + R)) = 2 × ((TP) / (TP + FP + FN)).

The ROC curve plots the probability that the model will correctly rank a randomly chosen positive instance higher than a random negative one [46], [48]: it represents the trade-off between recall and false positive rate, where the false positive rate is FPR = FP / (TP + FP).

We review in Table 1 the metrics utilized for evaluating the performance of detectors in recent conference papers focusing on Anomaly Detection (AD), either to build EDs or IDs. Our review includes publications since 2021 from the IEEE/IFIP International Conference on Dependable Systems and Networks (DSN), the IEEE International Symposium on Software Reliability Engineering (ISSRE) and USENIX Security Symposium. We delve into papers that explore datasets collected from various system features, obtained through diverse system monitoring tools. The review works are well-aligned on the most famous metrics previously reviewed. In addition, attention is given to the time required to analyze the input and prompt a decision [50], [51], [52].

In general, the related works aim to maximize the number of data points correctly classified. Aside Table 1, a significant exception is [5], where the concept of correct detection is bound to the concept of sequence of data points: a sequence of anomalous data points is correctly detected if at least one of the data points is identified as anomalous. A time-related approach is shown in [49], where the authors measure the size of the input windows required to detect an attack. This can be interpreted as an estimation of latency, even if it is bound to the investigation of the window size.

We argue that, depending on the operational context, the metrics reviewed may present a partial understanding of the actual capability of the detector. For example, while the accuracy seems satisfactory, the detector may have missed the initial part of the anomalous pattern, which still might be sufficient for an attacker to reach at least a partial goal.

Given that the temporal context is a key aspect of the anomaly detection problem, we argue that the time necessary to detect an anomalous pattern is important information to evaluate the real capabilities of a detector in an operational context. This detection time is traditionally described by the error or attack latency. We can define *error latency* as the temporal distance from the activation of an error and its detection [75] while *attack latency* is the time from the beginning of an attack until its detection.

There is also a practical reason, bound to the construction of datasets, to motivate the relevance of measuring latency when evaluating IDs and EDs. The fundamental observation is that, when an attack or error is first injected, its effect may manifest later in the monitored data. That is, we know the injection time and the corresponding data point, but we do not know the first data point that will have an observable anomalous behavior. In this perspective, the objective of correctly classifying all the data points logged during an attack, or after the activation of an error, is not realistic. The primary goal of a detector should be limited to identifying the attack promptly, or detecting an error before it escalates into a failure, obviously while minimizing false positives.

We better explain our position with an example. An attack may mix both totally legitimate operations and malevolent actions, like the sending of a forged network packet. The first will not trigger any anomalous behavior, but they are still part of the attack path. Consequently, the corresponding data points should be labeled as part of the attack sequence in the dataset. Additionally, the effect of the malevolent action may alter the feature value at a later point in time, possibly while legitimate actions are being executed. In this case, the detection task should focus on the detection of the attack sequence, rather than classifying individual data points. We believe this is a more realistic and representative scenario to study the ability to build IDs and EDs.

| Paper | Venue | Year | Scope | Metrics |
|---|---|---|---|---|
| Jha et al. [50] | DSN | 2022 | ED | P, R, F1, Lead Detection Time. |
| Wang et al. [10] | DSN | 2022 | ED | P, R, F1 |
| Dayaratne et al. [62] | DSN | 2022 | ID | P, R, F1, FPR |
| Alharthi et al. [63] | DSN | 2021 | ED | P, R, F1, MCC |
| Yuan et al. [64] | DSN | 2021 | ID | P, R, TPR, FPR |
| Xu et al. [69] | DSN | 2021 | ED | P, R, F1 |
| Zhao et al. [6] | DSN | 2019 | ID | A |
| Wang et al. [65] | ISSRE | 2022 | ED | P, R, F1 |
| Zhang et al. [66] | ISSRE | 2021 | ED | P, R, F1 |
| Jia et al. [67] | ISSRE | 2021 | ED | P, R |
| Zhang et al. [68] | ISSRE | 2021 | ED | P, R, F1, ROC |
| Alsaheel et al. [70] | USENIX | 2021 | ID | P, R F1, ROC |
| Chen et al. [51] | USENIX | 2021 | ED | R, avg. time |
| Downing et al. [71] | USENIX | 2021 | ID | P, R, FPR, ROC |
| Izhikevich et al. [52] | USENIX | 2021 | ED | A, proc. time |
| Fu et al. [72] | USENIX | 2021 | ED | P, R, FPR |
| Tang et al. [73] | USENIX | 2021 | ID | TPR, FPR |

Table 1: Review of metrics from recent conference papers on anomaly detection in the security and dependability domains. Only papers that deal with features collected from a system and processed as tabular data points are considered. Papers dealing with anomalies in images or text are left out.

## 3. Measuring Detection Latency

We define a *time series* as $X = \{\langle x_t \to x_{t+1}\rangle \mid x_t \in X, t \in [0, T]\}$, where $x_t$ is the data point collected at time $t$, and $x_{t+1}$ is the next data point after $x_t$, collected at time $t+1$, [53].

An AD dataset based on time series can exploit the temporal dimension by having multiple *sequences* of data points. A sequence concatenates data points collected while the system is operating in the absence of attacks/errors (we call it *normal operations*), and data points collected after an attack/error is injected (we call it *anomalous operations*). This reproduces the scenario of an attacker trying to penetrate the system, or of an error occurring in an error-free system. More formally, we define a *sequence* of length $N$ and injection instant $t_i$ as the concatenation of two time series $S_1$ and $S_2$:

$S_1 = \{\langle s_t \to s_{t+1}\rangle \mid s_t \in S, t \in [0, T_{i-1}]\}$
$S_2 = \{\langle s_t \to s_{t+1}\rangle \mid s_t \in S, t \in [T_i, T_N]\}$,

where data points $[t_0; t_{i-1}]$ are collected during normal operation of the system, and data points $[t_i; t_N]$ are collected during anomalous operations.

The detection of an attack or error in a sequence means correctly classifying a data point $x_t$ that belongs to the anomalous operations.

To present the metrics we will use in this paper, we need to formalize the definition of *attack and error latency* presented in Section 2.3 and from [75]. Given:
i) a sequence $S$, and an attack or error injected at time $t_i$, which marks the beginning of the anomalous operations,
ii) $s_i$ the data point of S collected at time $t_i$, and
iii) $x_d$ the value at time $t_d$ when detection is performed ($t_d \geq t_i$),

we define *latency* as $\Delta l = t_d - t_i$. In practice, we can measure latency as a time interval, or as the number of data points that occurred between the two data points $s_i$ and $s_d$.

In case the attack or error is not detected, the latency cannot be measured: this is the situation in which the system experiences a failure, or the attack exploit succeeds. For this reason, we compute the *sequence detection rate SDR*, expressed as the number of sequences where an attack or error is detected over the total number of sequences. In other words, $SDR = detected\ sequences\ /\ total\ sequences$. We can measure the *average latency* $\Delta L$ over the $N$ sequences, where $\Delta L = \sum_{i=0}^{n} \Delta l_i$.

Let us discuss some practical implications. A short latency can be easily achieved at the cost of an high FPR, which can be obtained setting a low detection threshold (used by the anomaly detectors to discern whether a data point is normal or anomalous). This would make the system unusable, because of too many and too frequent false alarms. The natural implication is that the measure of latency needs to be weighed against the FPR.

In an analogous way to how the precision-recall curve and the ROC are computed, we measure the *latency at different FPR thresholds*: this presents the tradeoff between latency and FPR. We expect that by increasing the FPR, the detection latency will be reduced because the detector will be more suspicious and tend to increase the data points classified as anomalous. The relationship between FPR and latency can be used to set the best trade-off based on the system requirements and the application domain of the detector.

The opposite (computing the FPR given a target latency) can also be measured, but it is not effective in explaining the performance of a detector. We justify our statement with the help of Figure 1. Let us suppose that we set a target latency $\Delta l_3 = t_3 - t_0$ According to Figure 1, all data points with confidence lower than 0.2 should be classified as normal, and those with confidence higher than 0.2 should be classified as anomalous. Using this information, the FPR can be computed easily. However, with such confidence threshold, the first anomalous data point correctly detected corresponds to $t_1$, so the actual latency would be $\Delta l_1 = t_1 - t_0$ which is smaller than $\Delta l_3$. This means that a lower latency $\Delta l_1$, an higher confidence threshold 0.3, and correspondingly a lower FPR better describe our detector.

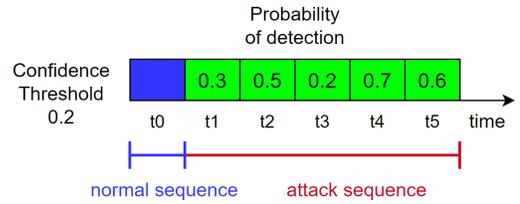

Figure 1: Relationship between latency values and confidence thresholds.

We remark that these metrics have practical value only if the dataset in use is composed of sequences of normal and anomalous operations as previously introduced. They are not useful when the dataset does not have these characteristics. For example, shuffling the data points is often necessary, due to the characteristics of the datasets [20], [26], [16], [28]: the realistic alternation of normal and anomalous behavior is compromised, and the measurement of latency would be unrealistic.

In the following of our paper, we show through an experimental campaign how we use this trade-off to study and calibrate a detector.

## 4. Experimental Campaign

### 4.1 Methodology

**M1.** We collect two public datasets: ROSpace [54], and Arancino [30]. The first is a dataset for intrusion detection composed by monitoring an embedded system on normal behavior and under attack. The second is specific for error detection and it represents an embedded system in an Internet of Things setting. Both datasets are time series-based and comprise a set of sequences (see Section 3) of variable length. The datasets are detailed in Section 4.2.

**M2.** We preprocess each dataset to extract sequences and shuffle their order. Note that we do not alter the order of data points within each sequence.

**M3.** We select 5 representative detectors from the state-of-the-art solutions reviewed in Section 2.2, organized into algorithms for unordered data points and algorithms for ordered data points. The algorithms are detailed in Section 4.3.

**M4.** We split the two datasets into train, validation, and test sets, using 60% of sequences for the training, 30% for the test, and 10% for validation. We train the selected detectors on each dataset.

**M5.** We evaluate the models using state-of-the-art metrics (accuracy, f1-score, precision, recall, precision-recall curve, ROC curve) and our proposed metrics latency, average latency, and sequence detection rate. We calculate attack latency in the case of ROSpace, while we compute error latency for the Arancino. The main criterium is that FPR should be a low value to reduce the number of false alarms. The metrics applied are summarized in Section 4.4.

**M6.** We discuss the results in Section 5.

### 4.2 Datasets

ROSpace [54] is a dataset for intrusion detection composed by performing penetration testing on SPaCe, an embedded cyber-physical system built over Robot Operating System 2 (ROS2 [55]). The SPaCe system is designed to orchestrate and optimize public mobility, improving the user experience and preventing security violations. It is implemented as an embedded cyber-physical system that relies on data collected by sensors to manage and determine the operative conditions of the vehicles in real time. The system analyses the collected data to establish the occupation of the vehicles and notify users and administrators of the system to optimize the availability of the service. It also provides detection capabilities to spot any dangerous situations like damage to vehicle equipment.

Features are monitored from three architectural layers: the Linux operating system, the network, and the ROS2 services. It collects a total of 6 attacks discovery and DoS attacks, with 3 of them specific to ROS2. The dataset is organized as a time series in which sequences of normal (attack-free) operations, and sequences when attacks are carried out in addition to the normal operations, are alternated. The goal of this strategy is to reproduce multiple scenarios of an attacker trying to penetrate the system. Noteworthy, this allows measuring the time to detect an attacker and the amount of malicious activities performed before detection. ROSpace includes 30 247 050 data points (78% attacks, 22% normal), and 482 columns excluding the label, for a total of 40.5 GB. We use the lightweight version of the ROSpace dataset that is reduced to the 61 most informative features (30 features from the Linux operating system, 5 from ROS2 services, and 25 from the network). Further, for computation constraints, we downsample the dataset by deleting 1 row each 3, obtaining a reduced dataset of 10 081 647 and 61 features (excluding label). The total number of sequences is 1173.

Arancino [30] is an embedded system for Internet of Things applications that is composed of two main parts: a microprocessor and a microcontroller. The dataset, presented in [30], provides data about how the Arancino device behaves in case of anomalies due to different injected errors. The system is monitored while executing normal operations and while injecting 8 errors. Once activated, the injection remains active for 5 seconds then the system waits for a total of 10 seconds of cooldown, in which memory is freed and no errors are injected. We assume that at the end of 5 seconds, the system fails, that is, we are required to detect errors before this period. The dataset is composed of 3 separate labeled datasets obtained by running the monitoring methodology in 3 different network environments [30]. The dataset counts 154 000 observations and 276 features other than the timestamp and a label, i.e., normal or any of the 8 errors. The total number of sequences 5273. After preliminary runs, we decided to augment the Arancino dataset to facilitate the detectability of its anomalous operations within each sequence. We replicate 3 times each anomalous data point $x_i$, and place the copied data point between the original $x_i$ and the subsequent data point $x_{i+1}$. Finally, we rearrange the timestamp accordingly to include the new data points, obtaining consecutive data points at 1-second intervals as in the original dataset.

The structure of the datasets is summarized in Table 2.

### 4.3 Detectors

We validate our methodology by running 5 different ML algorithms on the two datasets.

**Algorithms for unorder data points.** We use XGBoost, Extremely Randomized Trees, and Isolation Forest to detect punctual anomalies. XGBoost is a supervised decision-tree-based ensemble machine learning algorithm that uses a gradient boosting algorithm to a known dataset and then classifies the data accordingly [56], [18]. We use the code provided by [57]. We configure XGBoost with the following parameters: 10 gradient boosted trees, maximum tree depth for base learners set to 20, and learning rate of the boosting algorithm set to 1. We use the binary logistic loss metric.

Extremely Randomized Trees [58] is a random forest model based on ensembles. As in random forests, a random subset of candidate features is used, but thresholds are drawn at random for each candidate feature, and the best of these randomly generated thresholds is picked as the splitting rule. The training is performed in a supervised fashion. We use the implementation available at [59] and we configure the model to run with 100 estimators, which are the number of trees in the forest.

|  | attack/error | data points | sequence number | average duration | min. duration | max. duration | std. durat. | average length | min. length | Max. length | std. length |
|---|---|---|---|---|---|---|---|---|---|---|---|
| **ROSpace** | mflood | 4899066 | 294 | 60.709 | 59.000 | 60.344 | 8.233 | 43303 | 812 | 171799 | 22562 |
|  | ndisco | 166786 | 362 | 21.890 | 19.665 | 26.20 | 1.154 | 1377 | 1180 | 8248 | 232.80 |
|  | nflood | 2664621 | 20 | 29.848 | 29.658 | 29.960 | 0.079 | 398655 | 467044 | 40534 | 16319 |
|  | rosc | 1946 | 101 | 1.2415 | 1 | 40.232 | 6.817 | 51 | 2 | 150 | 20.477 |
|  | ros_rec | 133901 | 214 | 12.725 | 9.397 | 60.26 | 11.54 | 1907 | 1295 | 4873 | 251.54 |
|  | ros_ref | 1321 | 182 | 7.258 | 1 | 40.2 | 8.698 | 24 | 1 | 99 | 7.157 |
| **Arancino** | disks | 9396 | 619 | 14.18 | 14 | 17 | 0.71 | 15.18 | 15 | 18 | 0.71 |
|  | mems | 9279 | 581 | 14.97 | 14 | 17 | 1.4 | 15.97 | 15 | 18 | 1.4035 |
|  | rediss | 9174 | 604 | 14.18 | 14 | 17 | 0.73 | 15.18 | 15 | 18 | 0.73 |
|  | deadl | 8952 | 596 | 14.02 | 14 | 17 | 0.24 | 15.02 | 15 | 18 | 0.24 |
|  | cpus | 8805 | 582 | 14.13 | 14 | 17 | 0.6 | 15.13 | 15 | 18 | 0.6 |
|  | httpr | 8760 | 585 | 13.97 | 14 | 17 | 0.27 | 14.97 | 15 | 18 | 0.27 |
|  | proch | 8694 | 579 | 14.01 | 14 | 17 | 0.21 | 15.01 | 15 | 18 | 0.21 |
|  | redm | 8547 | 572 | 14.94 | 14 | 17 | 0.41 | 13.94 | 12 | 15 | 0.41 |
|  | proch | 8343 | 554 | 14.05 | 14 | 17 | 0.42 | 15.05 | 15 | 18 | 0.42 |

Table 2: Structure of the ROSpace and Arancino datasets. Columns indicate i) the attack and errors injected, ii) the number of data points (points num.), iii) the number of sequences (seq. num.), along with the corresponding average (avg. dur), iv) minimum (min.dur.), and maximum durations (max. dur.) in seconds, and their respective standard deviations (std.dur.), v) the average (avg. len.), minimum (min. len.), and maximum sequence lengths (max. len) in terms of number of data points, along with the associated standard deviations (std. len).

Isolation Forest [22] is an unsupervised algorithm constructed as an ensemble of Isolation Trees, where each tree organizes data points as nodes. The underlying assumption is that anomalies represent rare events with feature values significantly deviating from those of expected data points. Consequently, anomalies are more likely to be isolated closer to the root of the tree rather than the leaves. This logic implies that a data point can be effectively isolated and subsequently classified based on its distance from the root of the trees within the forest. We use the implementation available at [59]. We set for ROSpace a contamination parameter, indicating the assumed proportion of outliers in the dataset, to 0.25, which resembles the percentage of normal data points in the train set.

**Algorithms that exploit ordered data points.** We train an LSTM model which is composed of a convolutional neural network and two LSTM layers that process the input data points sequentially. Stacking multiple convolutional layers, the model learns hierarchical representations of the input, taking advantage of the order in which data points are provided. The lower layers capture low-level temporal patterns, while higher layers learn more abstract and complex patterns. The output of the convolutional network is passed as input to the LSTM model which can capture temporal dependencies from time ordered data points. We set *return_sequence = True* for the first LSTM layer to return the full output sequence, while we set it as False for the second, to output only the label prediction. The implementation can be found in [60]. The first layer is composed of 64 units, while the second of 32. We use SGD as optimizer with a learning rate configured to 0.001, decay equal to $10^{-8}$, and momentum set to 0.9.

Additionally, we train an improved model that combines machine learning techniques and LSTM technology for sequence classification tasks (LSTMCD) [61]. The model architecture consists of convolutional layers, followed by LSTM (Long Short-Term Memory) layers, and dense layers. The idea is to use the convolutional network to capture local patterns in the input sequences identifying features that are important for classification. In particular, we use two 1D convolutional layers (Conv1D) each one followed respectively by a batch normalization (BatchNormalization) and ReLU activation (ReLU). The first convolutional layer has 64 filters with a kernel size of 5, the second has 64 filters with a kernel size of 3. Each of the LSTM layers is configured with 64 units each. The output of the LSTM layers are fed to the two dense layers to improve the understanding of the sequences. We configure these layers to have 100 units, while the subsequent dense layer restricts the dimension to 50 units. The output is fed to the final dense output layer of two units, that use softmax to prompt out the final binary classification outcome.

To execute LSTM and LSTMCD, it is required to preprocess the train, test, and validation sets to obtain an input composed of 10 subsequent data points, obtained by selecting for each point $x_t$ the points between $x_t$ and $x_{t+9}$.

### 4.4 Metrics

We summarize in Table 3 the metrics we use for the evaluation of the selected models. In addition, we compute the ROC Curve and the precision-recall Curve which calculates the trade-off for different detection thresholds, respectively, between false positive rate and recall, and between precision and recall.

| Metrics | Formula |
|---|---|
| accuracy A | A = (TP+TN) / (TP+TN +FP+FN) |
| false positive rate FPR | FPR = FP / (TP + FP). |
| f1-score F1 | F1 = 2 × ((P × R) / (P + R)) |
| recall R | R = TP / (TP + FN) |
| precision P | P = TP / (TP + FP) |
| latency L | $\Delta l = x_t - x_d$ |
| average latency AL | $\Delta L = \sum_{i=0}^{n} \Delta l_i$ |
| sequences detection rate SDR | S = detected/total |

Table 3: The metrics we use for performance evaluation in our approach.

## 5. Experimental Campaign

### 5.1 Results with state-of-the-art metrics

First, we compute the ROC and precision-recall curves to have a preliminary understanding of the classification capabilities of the detector in different settings. In Figure 1 we show results for LSTMCD and XGBoost trained on both ROSpace and ARANCINO datasets. This shows that there is some detection capability, especially for XGBoost. The other algorithms are not reported for brevity but they have similar results.

Next, we study the amount of misclassification and if it can be considered acceptable for a realistic operation of the detector. We rely on Table 4. We set an upper limit of 0.01 to the FPR threshold for all the detectors. This is 1 false positive out of 100 data points. With such maximum FPR, the value of precision easily gets close to 1, except for cases labeled with *NaN* in the table, where no correct detections occurred. Given this consideration, precision is not shown in Table 4. The results for the Isolation Forest and LSTM on Arancino were not sufficiently good (near to random guessing): for this reason, results are not in Table 4.

XGBoost shows the best performances for both Arancino and ROSpace datasets, surpassing algorithms for unorder data points and for ordered data points.

The ROSpace attacks are best detected by XGBoost and LSTMCD. XGBoost achieves an accuracy of 0.927, a recall of 0.991, and an f1-score of 0.952. LSTMCD shows a slightly better performance in the f1-score with 0.953 while losing the comparison in terms of accuracy and recall, respectively 0.879 and 0.911. The LSTM model is the worst in terms of performance: recall and f1-score are near 0.

For the Arancino dataset, again the detectability is overall low. The best is XGBoost with an accuracy of 0.851, recall of 0.634, and f1-score of 0.769. Extremely Randomized Trees show similar results with an accuracy of 0.809, recall of 0.524, and f1-score of 0.681, while LSTMCD is much behind.

Both datasets are evidently difficult, meaning that the anomalous behavior of the system is not easy to observe with the available features. Also, the better performances of XGBoost suggest that the observable anomalies in both datasets are punctual rather than contextual or collective.

### 5.2 Latency-related metrics

In Table 5 we report the results obtained with the proposed evaluation methodology. We measure the sequence detection rate and the average latency. In the Table, we initially set an FPR threshold of 0.01.

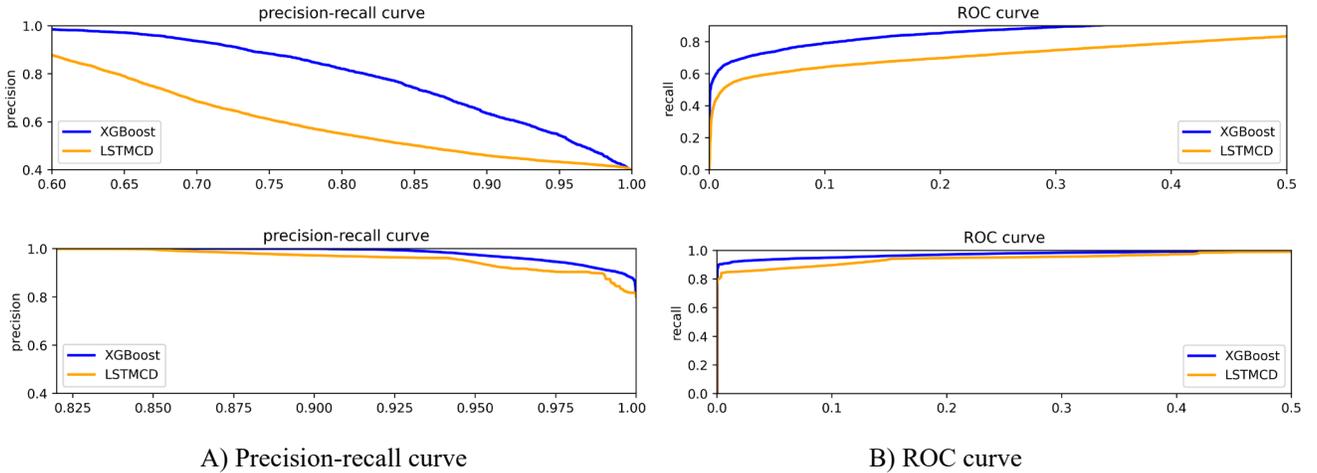

A) Precision-recall curve          B) ROC curve

Figure 2: Precision-recall and ROC Curves for XGBoost and LSTMCD on Arancino (top) and ROSpace (bottom).

| | Attacks/error | Detectors | | | | | | | | | | | | | |
|---|---|---|---|---|---|---|---|---|---|---|---|---|---|---|---|
| | | XGBoost | | | Extr. Random. Trees | | | Isolation Forest | | | LSTMCD | | | LSTM | | |
| | | A | R | F1 | A | R | F1 | A | R | F1 | A | R | F1 | A | R | F1 |
| ROSpace | m flood | 0.89 | 0.89 | 0.94 | 0.865 | 0.865 | 0.927 | 0.86 | 0.86 | 0.924 | 0.872 | 0.872 | 0.932 | 0.001 | 0.001 | 0.003 |
| | ndisco | 0.356 | 0.356 | 0.53 | 0.72 | 0.72 | 0.84 | NaN | NaN | NaN | 0.75 | 0.75 | 0.86 | NaN | NaN | NaN |
| | nflood | 0.999 | 0.999 | 0.999 | 0.09 | 0.09 | 0.15 | 0.018 | 0.018 | 0.037 | 0.856 | 0.856 | 0.922 | 0.03 | 0.03 | 0.07 |
| | rosc | NaN | NaN | NaN | NaN | NaN | NaN | NaN | NaN | NaN | NaN | NaN | NaN | NaN | NaN | NaN |
| | ros_rec | 0.007 | 0.007 | 0.014 | 0.002 | 0.002 | 0.004 | NaN | NaN | NaN | NaN | NaN | NaN | NaN | NaN | NaN |
| | ros_ref | NaN | NaN | NaN | NaN | NaN | NaN | NaN | NaN | NaN | NaN | NaN | NaN | NaN | NaN | NaN |
| | **all** | **0.927** | **0.991** | **0.952** | **0.61** | **0.52** | **0.683** | **0.57** | **0.477** | **0.644** | **0.879** | **0.911** | **0.953** | **0.20** | **0.001** | **0.003** |
| Arancino | disks | 0.986 | 0.986 | 0.993 | 0.956 | 0.956 | 0.977 | --- | --- | --- | 0.979 | 0.979 | 0.989 | --- | --- | --- |
| | mems | 0.773 | 0.773 | 0.872 | 0.781 | 0.781 | 0.877 | --- | --- | --- | 0.862 | 0.862 | 0.926 | --- | --- | --- |
| | rediss | 0.982 | 0.982 | 0.991 | 0.962 | 0.962 | 0.98 | --- | --- | --- | 0.98 | 0.98 | 0.99 | --- | --- | --- |
| | deadl | 0.313 | 0.313 | 0.477 | 0.241 | 0.241 | 0.388 | --- | --- | --- | 0.2 | 0.2 | 0.34 | --- | --- | --- |
| | cpus | 0.861 | 0.861 | 0.925 | 0.578 | 0.578 | 0.733 | --- | --- | --- | 0.54 | 0.54 | 0.70 | --- | --- | --- |
| | httpr | 0.507 | 0.507 | 0.673 | 0.256 | 0.256 | 0.407 | --- | --- | --- | 0.031 | 0.031 | 0.061 | --- | --- | --- |
| | proch | 0.129 | 0.129 | 0.229 | 0.15 | 0.15 | 0.26 | --- | --- | --- | 0.000 | 0.000 | 0.001 | --- | --- | --- |
| | redm | 0.959 | 0.959 | 0.979 | 0.573 | 0.573 | 0.729 | --- | --- | --- | 0.662 | 0.662 | 0.796 | --- | --- | --- |
| | proch | 0.132 | 0.132 | 0.233 | 0.133 | 0.133 | 0.235 | --- | --- | --- | 0.002 | 0.002 | 0.004 | --- | --- | --- |
| | **all** | **0.851** | **0.634** | **0.769** | **0.809** | **0.524** | **0.681** | --- | --- | --- | **0.794** | **0.487** | **0.648** | --- | --- | --- |

Table 4: Results obtained on ROSpace and Arancino. NaN means no correct detections at all, while the dash means the algorithm was random guessing.

XGBoost shows the best performances in terms of SDR, both on ROSpace and Arancino. In the first case, XGBoost and LSTMCD show similar performances with SDR of 0.50 and 0.494, which means there is at least one correct detection in half of the sequences.

Extremely Randomized Trees and LSTM are worse than XGBoost, with an SDR of 0.388 and 0.34, respectively: there is at least a correct detection in a third of the sequences. The worst results are achieved by Isolation Forest; this is expected, because Isolation Forest is an unsupervised algorithm, which is typically less performing than supervised ones when there are no unknowns. A similar trend is measured on Arancino, but the gap between XGBoost and LSMCD being more evident.

Instead, AL varies significantly, depending on the considered attack and error. This can be attributed to the variable effect of the attacks and errors on the target system: different attacks and errors alter the features' values in different ways, and their visible effects manifest at different time intervals since injection.

We now compare the results of Table 4, ranked according to f1, and those of Table 5, ranked according to SDR. In the case of Arancino, the ranking is the same: XGBoost shows the best overall performance, followed by Extremely Randomized Trees and LSTMCD. However, the gap between XGBoost and the other models in Table 4 is much more evident than in Table 5. In fact, in the case of Table 4, XGBoost detects many anomalous data points in the same sequence. If we are interested in the first detection of an anomaly, clearly Table 4 presents misleading outcomes. For example, ROSpace includes some sequences where the anomalous operations are composed of many data points, especially in the case of DoS attacks such as *nflood* (398 655 data points for each sequence of anomalous operations, see Table 2 and [74]). XGBoost correctly detects many of these data points, so its overall f1 is very good.

The most evident discrepancy between the two evaluation methods can be observed on the *deadl* error of the Arancino dataset. XGBoost obtains a recall of 0.313 and an f1 of 0.477, which is disappointingly low. The perspective about the detectability of such error is more optimistic considering SRD: XGBoost detects 84.4% of the *deadl* sequences.

Another interesting observation is with LSTM, which is generally showing poor classification performance. The *nflood* attack has a very low recall, but SDR=1.0. In this case, the model can recognize very few attack data points, but in all the sequences. This may suggest that there is a specific operation or packet sent by the attacker that always reveals the execution of an *ndisco* attack sequence, and the anomalies are detectable in only a part of the sequence.

Last, in Figure 2 we report on how the AL varies for different FPR. We consider only LSTMCD e XGBoost for brevity. As expected, increasing the FPR reduces the AL because the detector is more suspicious.

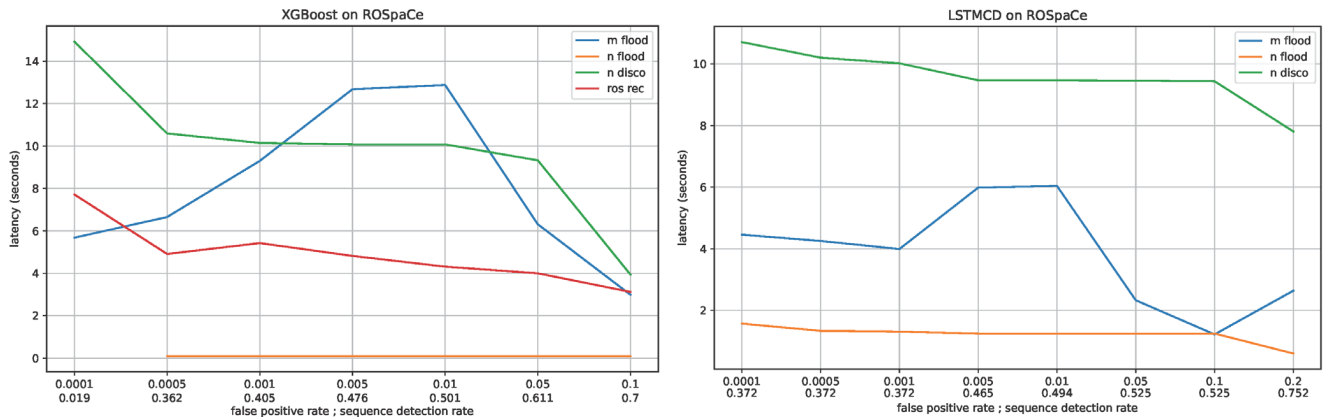

a) Average latency of XGBoost (left) and LSTMCD (right) on ROSpace, for different FPRs.

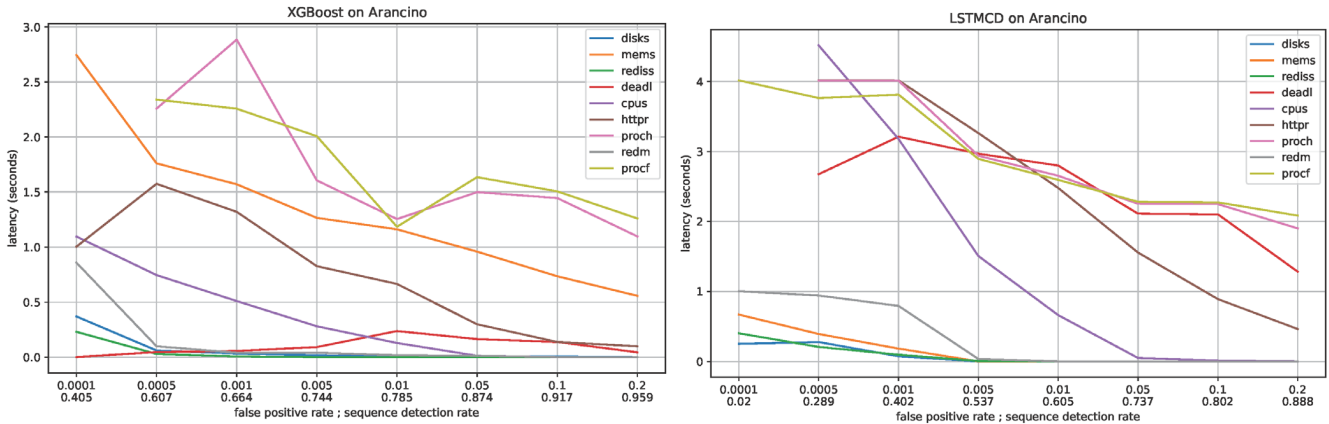

b) Average latency of XGBoost (left) and LSTMCD (right) on Arancino, for different FPRs.

Figure 3: Trade-off between AL and FPR for a) XGBoost and LSTMCD on ROSpace (top) and Arancino (bottom). FPR varies from 0.0001 to 0.2. SDR is also shown on the x-axis for completeness. *Better viewed in color.*

We can observe some spikes, like in the *mflood* attack and in Figure 2.a and the *disks*, *deadl*, *proch*, and *procf* in Figure 2.b. This is because, when the FPR raises, more sequences have at least a detection; such detections may be in a later part of the sequence, so the AL may increase. In other words, while the number of anomalous data points correctly detected does not increase significantly, these are distributed in more sequences. This contrasts with the general concept of the ROC curve or precision-recall curve that always show monotonic trends.

LSTMCD and XGBoost on ROSpace exhibit closely aligned AL across all attacks, with LSTMCD registering an AL that is generally shorter than XGBoost. This suggests that, on average, LSTMCD can detect attacks earlier than XGBoost for a given FPR.

In Arancino, LSTMCD and XGBoost have similar behavior, but the AL of XGBoost is shorter. This is probably because of two reasons: i) Arancino includes anomalies that can be classified as punctual, and they are placed in the initial part of the anomalous operations; ii) since LSTMCD uses a window of data points, it requires processing a larger piece of data than XGBoost before an anomaly is classified.

To summarize, Figure 2 shows a relationship between FPR and AL, while Table 5 shows the relationship between FPR and SDR- FPR can be tuned to set the desired AL and SDR, following operational requirements of the system. For example, in ROSpace, LSTMCD should be selected if the main desideratum is to detect flooding attacks with low FPR, or when a short detection time is fundamental. XGBoost should be preferred if the objective is to have a high SDR, i.e., increase the chance to detect an attack, even if at a later time.

## 6. Concluding remarks

Our evaluation methodology offers a framework for assessing the performance of intrusion and error detection models. Considering latency, false positive rate, and sequence detection rate, we can make decisions tailored to specific system requirements and operational priorities. Our approach provides valuable information about the effectiveness of different models, allowing for the selection of the most suitable detection strategy to preserve system security and resilience. Additionally, our methodology prevents biases introduced by dataset labeling methods, ensuring a realistic evaluation.

### 6.1 Position statement and main takeaway

We believe that a realistic assessment of IDs and EDs should be centered around detection latency and the ability to detect a sequence of anomalous events. In fact, in practice, we are usually interested in detecting one relevant event that will trigger response strategies that mitigate the adversarial actions. The rest of the adversarial actions may also be ignored once the attack path has been broken.

To realize this, we believe it is necessary to include the false positive rate, such that different detection latencies and the ability to detect anomalies can be related to false alarms. False alarms are usually a fundamental requirement for the deployment of IDs and EDs, as an excessive number of false alarms will result in useless detectors.

We have experimentally demonstrated that a clear relation between false positive rate, detection latency, and ability to detect a sequence of anomalous events exists. Despite the relation not being monotonic, in general, increasing the FPR would lead to a decrease in latency since the detector becomes more cautious and suspicious. We provided means to compute the above metrics, through the code at our repository [11].

However, we are aware that most of the public datasets available are not suitable for computing these metrics [74]. They are rarely structured in sequences of normal and anomalous behavior, and they usually label as anomalous a data point collected while an attack or error is injected, without evidence that the corresponding effect of the attack or error manifests in that specific data point and are thus observable by an algorithm. Hypothetically, the visible effect could manifest later. This requires that a change of position is also proposed when building datasets, such that the proposed methodology can be exploited.

### 6.2 Procedural steps

The application of our methodology can be summarized in the following recommended steps.

1. Organize the train and test sets in sequences, where each sequence is composed of normal operations followed by anomalous operations. Each sequence describes the system dynamics when an attack is launched or an error activates. Necessarily, the dataset must be labelled.
2. After training the detector, assess its performance using traditional state-of-the-art metrics.
3. Complement traditional metrics with AL and SDR measured at different FPRs. To implement this, it is just required to keep track of some additional information: i) the initial data point of each sequence, ii) the position of the first data point labelled as anomalous, and ii) the position of the first correctly classified anomalous data point in each sequence. Given this information, computing AL and SDR is straightforward.
4. Choose the target FPR threshold that allows matching system operational requirements. It is necessary to recompute traditional state-of-the-art metrics once the target FPR is decided.